\documentclass[preprint,a4paper,aps,pre,showpacs,showkeys,floatfix]{revtex4}

\usepackage{amsmath} \usepackage{amsfonts} \usepackage{amssymb}
\usepackage{graphicx}
\usepackage{dcolumn}
\usepackage{bm}

\newcommand{\figwidth}{0.95\columnwidth}

\begin{document}


\title{Aharonov-Bohm effect for an exciton in a finite width nano-ring}
\author{F.\ Palmero}
\email[Corresponding author. Electronic address: ]{palmero@us.es}
\affiliation{Grupo de F\'{\i}sica No Lineal. Departamento de
F\'{\i}sica Aplicada I., ETSI Inform\'atica. Universidad de
Sevilla, Avd. Reina Mercedes s/n, 41012-Sevilla, Spain}
\author{J.\ Dorignac}
\altaffiliation[Current address: ]{College of Engineering, Boston
  University, 44 Cummington Street, Boston, MA 02215}
\author{J.C.\ Eilbeck}
%
\affiliation{Department of Mathematics, Heriot-Watt University,
Riccarton, Edinburgh, EH14 4AS, UK}
\author{R.A.\ R\"{o}mer}
\affiliation{Department of Physics and Centre for Scientific Computing,
   University of Warwick, Coventry, CV4 7AL, UK}

\date{\today}

\date{$Revision: 2.4 $, compiled \today}

\begin{abstract}
  We study the Aharonov-Bohm effect for an exciton on a nano-ring
  using a 2D attractive fermionic Hubbard model. We extend previous
  results obtained for a 1D ring in which only azimuthal motion is
  considered, to a more general case of 2D annular lattices. In
  general, we show that the existence of the localization effect,
  increased by the nonlinearity, makes the phenomenon in the 2D system
  similar to the 1D case. However, the introduction of radial motion
  introduces extra frequencies, different from the original isolated
  frequency corresponding to the excitonic Aharonov-Bohm oscillations.
  If the circumference of the system becomes large enough, the
  Aharonov-Bohm effect is suppressed.
\end{abstract}



\maketitle


\section{Introduction}
\label{sec-introduction}

Progress in micro-structuring technology allows the construction
of nano-sized InGaAs rings by self-assembly
\cite{EmpPBL99,LorLFK99,LorLGK00} or using lithographic techniques
\cite{BaySHF00,BayGHM03}. Similarly, PbSe-based nano-rectangles
have been synthesized in solution through oriented attachment of
nanocrystal building blocks \cite{ChoTGM05}.
In these systems, electrons and holes can propagate coherently and the
existence of bound states --- so-called excitons --- offers an
opportunity to explore the Aharonov-Bohm (AB) effect
\cite{AhaB59,ByeY61}. The exciton, although neutral, can be sensitive
to a magnetic flux due to its finite sized Bohr radius \cite{RomR00}.
This effect, in the framework of excitons in nano-rings, has been
studied previously by means of simple models, such as a continuous
one-dimensional (1D) quantum ring
\cite{RomR00,Cha95,ChaG98,MasMTK01,MeiTK01}, or in continuous
two-dimensional (2D) models for which some numerical solutions can be
found \cite{HuLZX00}. The effects of an additional external electric
field have been also been studied \cite{MasC03}. In all cases, it has
been shown that AB oscillations can exist in a finite-width
nano-ring, but when the ring becomes large enough, the effect is
suppressed. A similar effect has also been predicted for bi-excitons
\cite{MeiTK01}.

In this paper we consider lattice models of such nano-rings with
an electron and hole subject to a perpendicular and uniform magnetic
flux. Our models allow for a finite number of transport channels,
either arranged annularly, i.e., coupled rings laid out in a 2D
plane with increasing radius, or stacked vertically. Therefore, we
can study in a controlled way how the excitonic AB oscillations
evolve when additional rings are added, either in-plane or
vertically.
We describe the electron-hole system on the rings by means of an
attractive Hubbard model where electron and hole are modelled as
spin-up and spin-down particles with different effective masses. We
focus our interest on the study of the AB oscillations, their
dependence on the size of the rings, and on different combinations of
lattice parameters.

For attractive Hubbard models, the existence of bound states of
electrons and holes --- identical to spin-singlets when using the
standard electronic interpretation --- is well known \cite{LieW68}, and
has been previously also been studied in the framework of quantum
breathers \cite{sc99}. In these cases, if the anharmonicity (interaction
strength) is strong enough, there is an extended Bloch state with two or
more particles in a strongly correlated state.  There exist various
theoretical \cite{seg94,Qtrev,Dor04} and numerical \cite{Eil04} results
as well as experimental observations of these states in different
quantum systems \cite{Fil,Asa00,Sch02}.

Our work is organized as follows: in the next section we present the
model. In Section \ref{sec-one_ring_case} we study the 1D model and
analytical expressions are obtained and compared to the 1D continuum
approach. In Section \ref{sec-two_ring_case}, we consider a simple
{\em plate-shaped} 2-ring model and compare the AB oscillations with
those obtained in the 1D case. In Section \ref{sec-general_case}, we
extend our results to a finite number ($>2$) of rings. In Section
\ref{sec-stack}, we consider a model in which the rings have been
stacked vertically, so that the same flux is threading each ring.
Finally, in Section \ref{sec-conclusions}, we summarize our findings
and present our conclusions.

\section{The model}
\label{sec-model}

\begin{figure}[tbh]
\begin{center}
 \includegraphics[width=\figwidth]{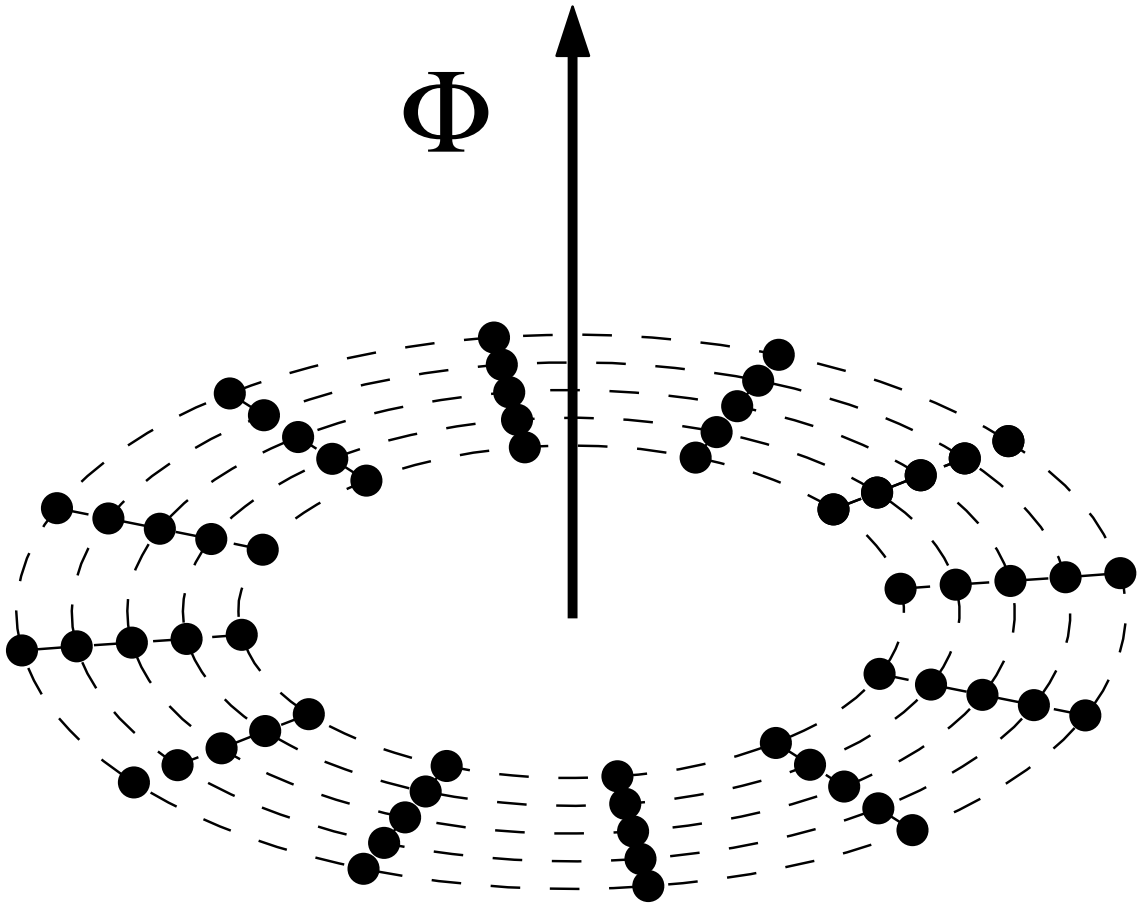}
 \end{center}
\caption{Sketch of the model. The lattice is a 2D system of ${N}$
rings and ${M}$ sites in each ring. The system is subject to an
uniform magnetic field perpendicular to the lattice, with $\Phi$
the magnetic flux.} \label{model}
\end{figure}

We consider an anharmonic 2D lattice of ${N}$ rings, each ring with ${M}$
sites, subject to a uniform magnetic field
perpendicular to
the lattice, as shown in Fig.\ \ref{model}.
On this lattice we consider an electron and a hole. We assume that
the interaction potential is short-ranged and the system can be
described by an attractive fermionic Hubbard model. Also, we will
consider only the possibility of hopping between nearest-neighbour
sites. Thus, the Hamiltonian is given by
\begin{align}
\hat H & =  -\gamma \sum_{n=1}^{{N}} \sum_{m=1}^{{M}} a_{n,m}^\dag
a_{n,m} b_{n,m}^\dag b_{n,m}-  \label{Ham_fer}\\
&  \sum_{n=1}^{{N}-1} \sum_{m=1}^{{M}} t^{\perp}_n \left[ a_{n,m}^\dag
a_{n+1,m}+a_{n+1,m}^\dag a_{n,m}+ \right. \nonumber \\ &  \left. \mu
\left( b_{n,m}^\dag
b_{n+1,m}+b_{n+1,m}^\dag b_{n,m}\right) \right] - \nonumber \\
&  \sum_{n=1}^{{N}} \sum_{m=1}^{{M}} t^{||}_n  \left[ e^{2 \pi i \varphi_n/{M}}
\left( a_{n,m}^\dag a_{n,m+1}+\mu
b_{n,m+1}^\dag b_{n,m}\right)+ \right. \nonumber \\
& \left. e^{-2 \pi i \varphi_n/{M}}
\left( a_{n,m+1}^\dag a_{n,m}+\mu b_{n,m}^\dag b_{n,m+1} \right)\right] \nonumber,
\end{align}
here $a_{n,m}^\dag (a_{n,m})$ and $b_{n,m}^\dag (b_{n,m})$ are
raising (lowering) operators for electrons and holes,
respectively, satisfying the standard fermionic anti-commutation
relations. The parameters $t^{||}_n$ represent the hopping
coefficient between neighbouring sites along the $n\,$th ring, and
$t^{\perp}_n$ the hopping coefficients between neighbouring sites,
one corresponding to the $n$th ring and the other one to the
$(n+1)$th ring (in absence of magnetic field). In general, these
coefficients will depend on the distance between sites. For an
annular ring structure as in Fig.\ \ref{model}, the site-to-site
distances increase as one moves further away from the centre. In
an single ring with no interaction, ($\gamma=0$), the effective
mass of the electron and hole at the extrema of the conduction and
valence band are proportional to the inverse of the square of the
distance between sites [12]. In our model, we have chosen, for a
fixed value value of the number of sites, a fixed effective mass
for all the rings with different radius and, therefore, we model
the intra--ring nearest--neighbour hopping coefficient  by
assuming a dependence according to an inverse-square law. Also, we
consider a similar dependence for the inter--ring hopping
coefficient. Thus,
\begin{equation}
t^{\perp}_n = \frac{\epsilon}{(r_{n+1}-r_n)^2}, \quad t^{||}_n        =
\frac{\epsilon}{4 r_n^2\sin^2(\pi/{M})},
\end{equation}
where $r_n$ is the radius of the $n$th ring and $d_n=2 r_n
\sin(\pi/{M})$ is the chord distance between two neighbouring lattice
points of the $n$th ring. Also, the formal continuum limit can be
obtained by taking this distance going to zero. We will suppose that the
distance between rings is constant, $t^{\perp}_n=t^{\perp}$, and thus
independent of the ring.  The parameter $\gamma/\epsilon $ represents
the ratios of anharmonicity (interaction parameter) to nearest-neighbour
hopping energy, $\mu$ the ratio of the effective masses of electrons and
holes at the bottom and the top of the conduction and valence band
respectively, and $\varphi_n=\Phi_n/\Phi_0$, $\Phi_n$ the magnetic flux
through the $n$th ring and $\Phi_0$ the flux quantum. In general we
consider $\epsilon=1$ and $\mu=0.2$, as in Ref.\ \cite{HuLZX00}.

The 1D Hubbard model with nearest-neighbour hopping terms has been
solved exactly in the seminal papers by Lieb and Wu \cite{LieW68}
and with flux by Shastry and Sutherland \cite{ShaS90,SutS90}.
Recent results are reviewed in Ref.\ \cite{EssFGK05}. Similarly
exact results for extensions of the model are rare. An
approximation to calculate the spectrum of the Hubbard model has
been proposed, see, e.g.\ \cite{GebR92}, where the exact spectrum
and thermodynamics for a long-range hopping Hubbard chain with
linear dispersion is calculated. In this paper, we will use the
number-state method  \cite{sc99} to calculate the eigenvalues and
eigenvectors of the Hamiltonian operator (\ref{Ham_fer}) in the
sector with a single electron and a single hole. We use a
number-state basis
\begin{eqnarray}
\lefteqn{|\psi_{\{n\}}\rangle  = } & &  \nonumber \\
& &
\left|
\begin{array}{l}
n^e_1,n^e_2,\ldots ,n^e_{M};n^e_{{M}+1},n^e_{{M}+2},\ldots , n^e_{2{M}};
n^e_{{M}+1}, \ldots
 \\
n^h_1,n^h_2,\ldots ,n^h_{M},n^h_{{M}+1},n^h_{{M}+2},\ldots , n^h_{2{M}};
n^h_{{M}+1}, \ldots
\end{array} \right. \nonumber \\
& & \mbox{ }\left.
\begin{array}{l}
\ldots; n^e_{{M}({N}-1)+1}, n^e_{{M}({N}-1)+2},\ldots ,n^e_{{M}{N}} \\
\ldots; n^h_{{M}({N}-1)+1}, n^h_{{M}({N}-1)+2},\ldots ,n^h_{{M}{N}}
\end{array} \right\rangle,
\end{eqnarray}
where $n^e_{(m-1){N}+n}$ ($n^h_{(m-1){N}+n}$) represents the
numbers of electrons (holes) at site $n$ of the ring $m$. We
always have $\sum_{n,m} n^e_{(m-1){N}+n}=1$ and $\sum_{n,m}
n^h_{(m-1){N}+n}=1$. A general wave function is
$|\Psi\rangle=\sum_{\{n\}} c_{\{n\}}|\psi_{\{n\}}\rangle$. It is
possible to block-diagonalize the Hamiltonian operator using
eigenfunctions of the periodic translation (or rotation) operator
$\hat T$ defined as $\hat T b_{n,m}^\dag = b_{n,m+1}^\dag \hat T$
($\hat T a_{n,m}^\dag = a_{n,m+1}^\dag \hat T$).  In each block,
the eigenfunctions have a fixed value of the {\em total} momentum
$K$, with $\tau=\exp(i K)$ being an eigenvalue of $\hat T$ such
that $K=2 \ell\pi/{M}$ and $\ell$ integer \cite{sc99}.

\begin{figure}[tbh]
\begin{center}
 \includegraphics[width=\figwidth]{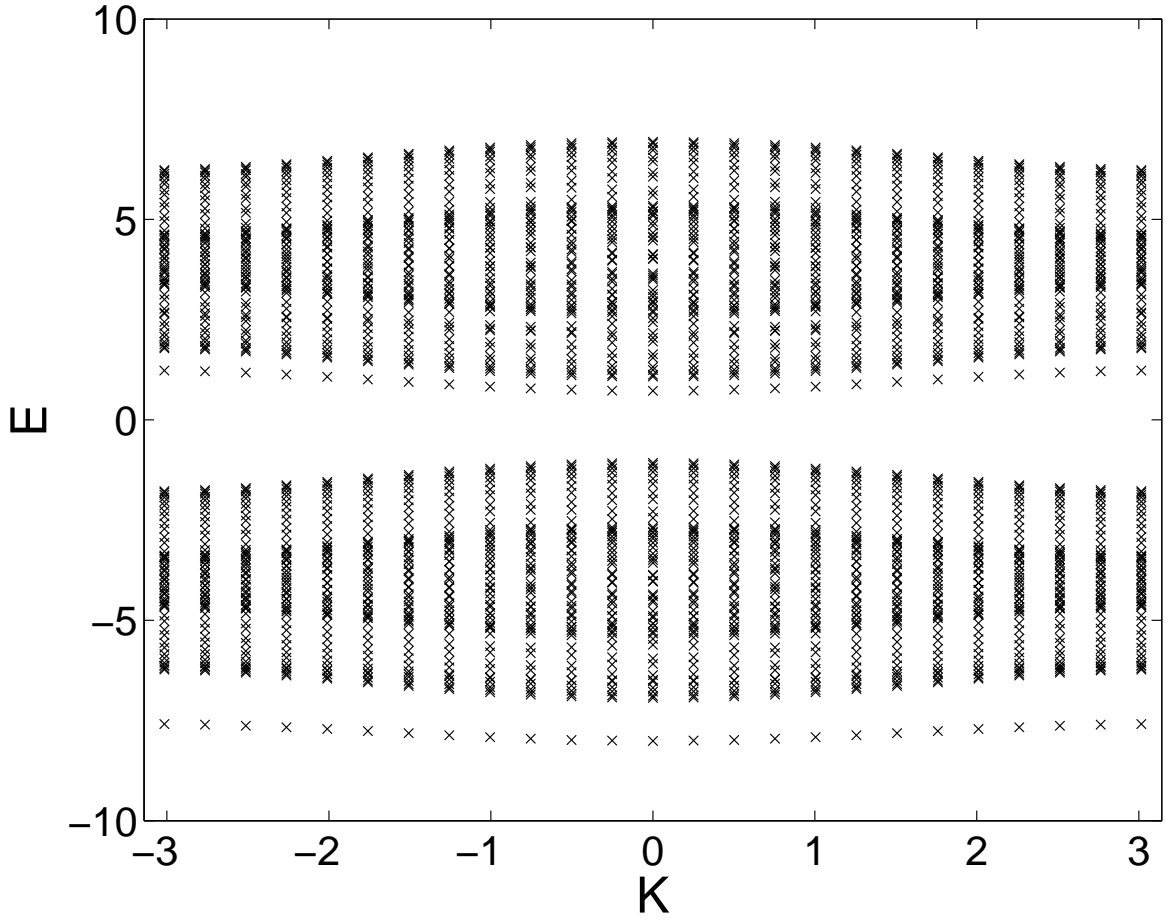}
  \caption{Band structure of the eigenvalues $E(k)$. ${M}=25$, ${N}=2$,
    $\gamma=4$, $\varphi_1=\varphi_2=0$, $d_1=1$, $\epsilon=1$,
    $\mu=0.2$.  $\Delta r=r_2-r_1=0.5$.}
  \label{band}
\end{center}
\end{figure}

In general, if the anharmonicity parameter is large enough, there
exists some isolated eigenvalue for each $k$ which corresponds to a
localized eigenfunction.  By ``localized'', we mean there is a high
probability for finding the two fermions at the same site. This is
illustrated by the band structure in Fig.\ \ref{band}, which shows two
isolated bands below the two ``continuum'' bands.  Due to the
rotational invariance of the system, there is the same probability of
finding two fermions at any site of a ring. In these cases, analytical
expressions can be obtained in some asymptotic limits
\cite{seg94,sc99,Eil03}.

\section{A single ring}
\label{sec-one_ring_case}

\subsection{The Hamiltonian matrix}
\label{sec-1d-intro}

In the simplest case, $N=1$, the system reduces to a 1D ring of ${M}$
sites and radius $r$. The Hamiltonian matrix in each momentum subspace
is given as
\begin{equation}
\label{ham_one} H^{(K)}_{1} = -\left[
\begin{array}{cccccc}
\gamma & q^* & 0 & . & . & q \\
q & 0 & q^* & 0& . & 0 \\
0 & q & 0 & q^* & . & . \\
. & . & . & . & . & . \\
. & . & . & q & 0 & q^* \\
q^* & . & . & . & q & 0 \end{array} \right],
\end{equation}
where $q=t^{||} e^{2 \pi i \varphi/{M}}\left(\mu+e^{-i K}\right)$ and
we have omitted the index labelling the ring number.
In the case of no interaction ($\gamma=0$) between the electron and
hole, the energy of the system is given by
\begin{equation}
E_{k^{e},k^{h}}=-2 t^{||} \cos\left(k^{e} -
\frac{2 \pi \Phi}{{M} \Phi_0} \right)-2 t^{||}
\mu \cos\left(k^{h} + \frac{2 \pi \Phi} {{M} \Phi_0}\right),
\label{armonic}
\end{equation}
where $k^{e}, k^{h}= 2 \pi \ell/M$ are the {\em single particle}
wave vectors in the absence of a magnetic field \cite{MasMTK01},
and $K= k^{e} + k^{h}$. In the continuous limit, when $M
\rightarrow \infty$, $d \rightarrow 0$ but the radius of the ring
is fixed, and considering the energy of the ground state in the
absence of the field for the electron and hole as zero, Eq.\
(\ref{armonic}) reduces to the standard parabolic expressions for
charges in a continuous ring threaded by a magnetic flux and no
interaction \cite{RomR00}.

\subsection{Exact results}
\label{sec-1d-exact}

For $\gamma \neq 0$, the ground state in each momentum sector
associated with (\ref{ham_one}) is given by \cite{sc99}
\begin{align}
|\Psi_{\tau}\rangle & =
c_0 \sum_{j=1}^{{M}} (\hat T/\tau)^{j-1}
\left|\begin{array}{l}
10\ldots 0\\
10\ldots 0
\end{array}\right\rangle\\
& \mbox{ }+ c_1 \sum_{j=1}^{{M}} (\hat T/\tau)^{j-1}
\left|\begin{array}{l}
10\ldots 0\\
01\ldots 0
\end{array}\right\rangle\\
&  \mbox{ } +\ldots +c_{{M}-1} \sum_{j=1}^{{M}} (\hat
T/\tau)^{j-1}
\left|\begin{array}{l}
10\ldots 0\\
00\ldots 1
\end{array}\right\rangle,
\end{align}
where
\begin{equation}
c_n=\frac{c_0 e^{i n
\theta}}{\sinh[{M}\nu]}\left\{{\sinh[({M}-n)\nu]+e^{-i
{M}\theta}\sinh(n\nu)}\right\},
\end{equation}
$\theta=\arg(q)$, $\cosh(\nu)=-E/2|q|$ and $c_0$ a normalization factor.
Details of the calculation can be found in appendix \ref{sec-eveceval}.
The energy spectrum $E_{\nu}=-2 |q| \cosh(\nu)$ is determined by
the solutions of the equation
\begin{equation}
\label{enb}
 \frac{\tanh({M}\nu)}{\sinh(\nu)}=\frac{2|q|}{\gamma}
\left[1-\frac{\cos({M}\theta)}{\cosh({M}\nu)}\right].
\end{equation}
As we show in appendix \ref{sec-bound}, bound state solutions to
(\ref{enb}) exist iff
\begin{equation}
\gamma > \frac{2 |q|}{M}(1-\cos M\theta).
\label{conditionb}
\end{equation}
These solutions correspond to the bound electron-hole pair, namely the
{\em exciton}.
In the limit ${M} \rightarrow \infty$, $d$ constant, the energy of the
exciton is $E_\infty=-\sqrt{\gamma^2+4|q|^2}$, independent of the value
of the magnetic field.  This value corresponds to the exciton binding
energy in a straight wire \cite{RomR00}.  Also, the
anharmonic/interaction parameter $\gamma$ must scale as $1/d$ in order
to support a bound state solution; it goes to infinity as
expected for a point-like $\delta$-function interaction
\cite{GE99}.

\subsection{Approximations for a large ring}
\label{sec-1d-large_ring}

If $M d$ is large, we can derive some explicit approximate
expressions (see appendix \ref{sec-largeM} for details). First,
the energy of the bound state is given by
\begin{equation} \label{EBS}
E = E_\infty\left[1 +\frac{2\gamma^2} {E_{\infty}^2} \cos (M
\theta)\, e^{-M {\rm arcsinh} \left( \frac{\gamma}{2|q|}
\right)}\right].
\end{equation}
This formula is close to the perturbative exciton energy obtained
for a large ring radius in the continuum model \cite{RomR00}. Also,
if $d$ is small enough, condition (\ref{conditionb}) can be
approximated as $2 \pi r \beta >1-\cos(2\pi\Phi/\Phi_0)$, where
$\beta={\rm arcsinh} ( \gamma /2|q|)/d$ is the inverse decay
length of energy oscillations given by Eq.\ (\ref{EBS}).

\subsection{Numerical results}
\label{sec-1d-numerics}

\begin{figure}[tbh]
\begin{center}
 \includegraphics[width=\figwidth]{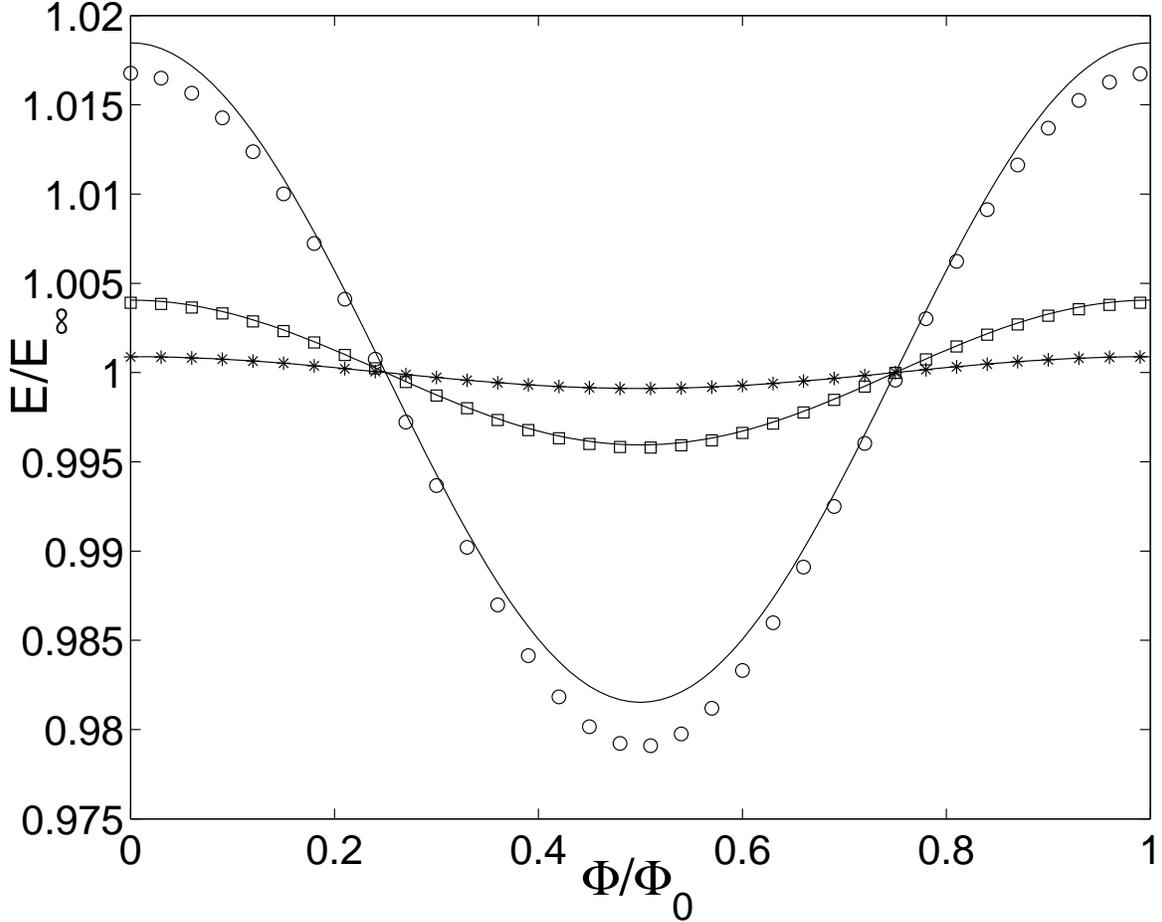}
 \caption{The Aharonov-Bohm oscillations of exciton energy (in units
   of $E_\infty$) corresponding to three values of the (single) ring
   circumference $\rho$ (in units of $d$) obtained by solving Eq.\
   (\ref{enb}) numerically for $\gamma=2$, $d=1$, $\epsilon=1$,
   $\mu=0.2$, and $k=0$. ($\circ$) ${M}=5$ ($\rho=5.34$), ($\square$)
   ${M}=7$ ($\rho=7.24$) and ($\ast$) ${M}=9$ ($\rho=9.19$). The lines
   denote the approximate solutions given by Eq.\ (\ref{EBS}).}
  \label{AB_o}
\end{center}
\end{figure}

In Fig.\ \ref{AB_o} we plot the energy of the ground state for several
values of the magnetic flux and the circumference $\rho=2 \pi r$ of
the ring. It is easy to see that the magnitude of the AB effect
decreases rapidly with increasing ring radius.  As expected
\cite{RomR00}, the maximum amplitude of the AB oscillations
corresponds to integer or half-integer values of $\Phi/\Phi_0$. If we
analyze the (unbound) excited states, the contributions of components
corresponding to states where fermions are localized in different
sites of the lattice are significant, and the AB contribution no
longer decrease exponentially with the circumference of the ring.

\begin{figure}[tbh]
\begin{center}
 \includegraphics[width=\figwidth]{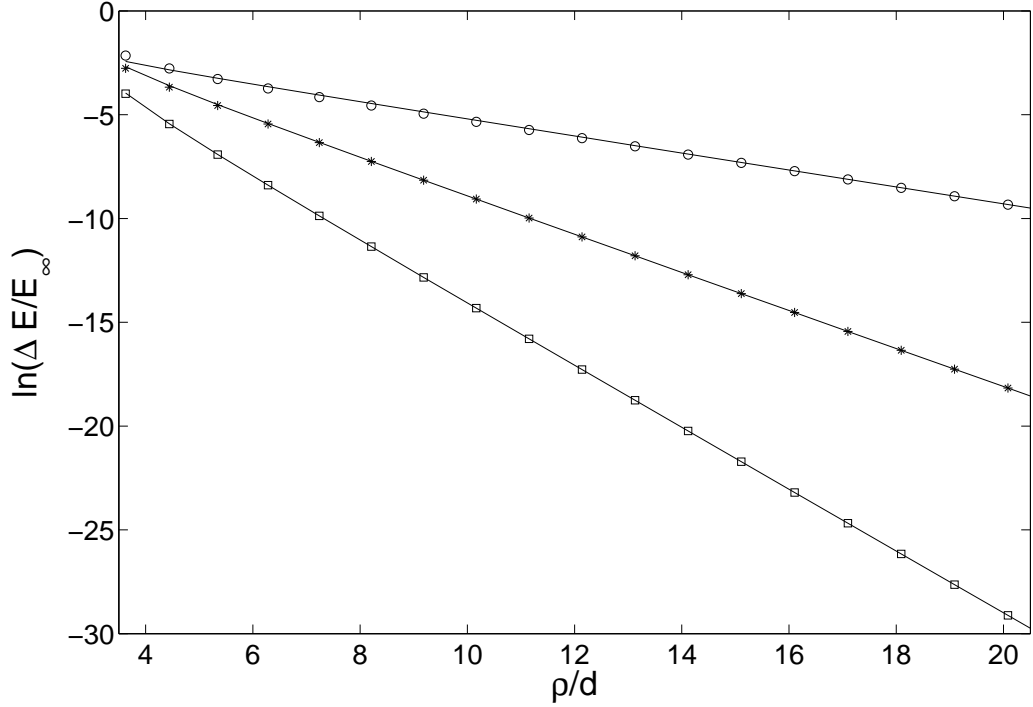}
 \caption{Energy oscillations amplitude of AB oscillations (in units
   of $E_\infty$ and on a semi--logarithm scale) as a function of the
   circumference of the (single) ring $\rho$ (in units of
   circumferential lattice spacing $d$), obtained by solving Eq.\
   (\ref{enb}) numerically for $\gamma=1$, corresponding to a weakly
   bounded state ($\circ$), $\gamma=2.5$, a typical bound state
   ($\ast$), and $\gamma=5$, corresponding to a strongly bound state
   ($\square$). $\epsilon=1$, $\mu=0.2$, and $k=0$.  The continuous
   lines are approximate solutions given by Eq.\ (\ref{EBS}). All logarithms
   are natural logarithms.}
  \label{energy_o}
\end{center}
\end{figure}

\begin{figure}[tbh]
\begin{center}
 \includegraphics[width=\figwidth]{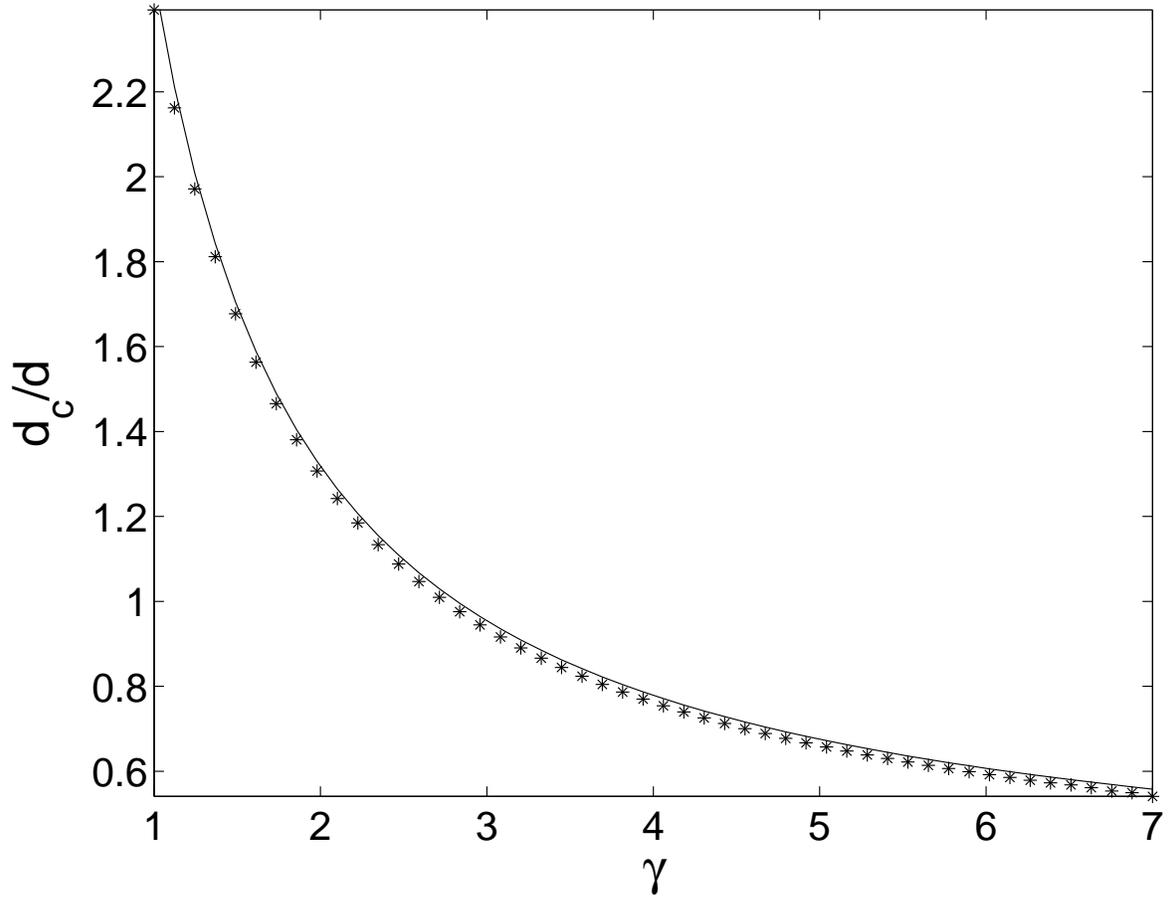}
 \caption{Decay length $d_c$ of AB oscillations (in units of the
   circumferential lattice spacing $d$) for a single ring as a function
   of the anharmonic parameter $\gamma$. $\epsilon=1$, $\mu=0.2$, and
   $k=0$, obtained by numerical solution of Eq.\ (\ref{enb}) ($\ast$).
   The continuous line is the approximate solution given by Eq.\
   (\ref{EBS}). }
  \label{decay}
\end{center}
\end{figure}

For the ground state, if we plot the amplitude of the AB
oscillations $\Delta E= E(\varphi=0)-E(\varphi=1/4)$ as a function
of circumference $\rho$ as shown in Fig.\ \ref{energy_o}, we
observe an exponential decay with a decay length $d_c$ which can
be calculated numerically by solving Eq.\ (\ref{enb}).
The characteristic decay length is a function of the anharmonic
parameter $\gamma$ and, as shown in Fig.\ \ref{decay}, it decreases
with increasing $\gamma$.

\section{The case of two rings}
\label{sec-two_ring_case}

\subsection{The Hamiltonian matrix}
\label{sec-2d-hamiltonian}

In the case of two rings ($N=2$) with $M$ sites on each, it is again
possible to find the Hamiltonian matrix $H^{(K)}$ in each total
momentum $K$ subspace,
\begin{equation}
\label{HK}
H^{(K)}_{2} = -\left[
\begin{array}{cccc}
 H_{11} & H_{12} & H_{13} &  H_{14} \\
 H^*_{12} & H_{22} & H_{23} &  H_{13} \\
 H^*_{13} & H^*_{23} & H_{33} & H_{12} \\
 H^*_{14} & H^*_{13} & H^*_{12} &  H_{44},
\end{array} \right]
\end{equation}
where each $H_{ij}$ submatrix is of dimension $ {M} \times {M} $.
The diagonal blocks are given by $H_{11}= {\cal F}(\gamma, q_1)$,
$H_{44}={\cal F}(\gamma,q_2)$, $H_{22}={\cal F}(0,q_{12})$, $H_{33}=
{\cal F}(0,q_{21})$ with the ${M}\times{M}$ matrix ${\cal F}(p,q)$
given by
\begin{equation}
{\cal F}(p,q) = \left[
\begin{array}{cccccc}
p & q^* & 0 & . & . & q \\
q & 0 & q^* & 0& . & 0 \\
0 & q & 0 & q^* & . & . \\
. & . & . & . & . & . \\
. & . & . & q & 0 & q^* \\
q^* & . & . & . & q & 0 \end{array} \right].
\end{equation}
The off-diagonal blocks are $H_{12}=\mu t^{\perp}_1 \openone$, $H_{13}=
t^{\perp}_1 \openone$, where $\openone$ is a ${M}\times{M}$ unit matrix,
%
and the parameters are
\begin{eqnarray}
q_1 & = &t^{||}_1 e^{2 \pi i \varphi_1/{M}}(\mu+e^{-i K}), \\
q_2 & = &t^{||}_2 e^{2 \pi i \varphi_2/{M}}(\mu+e^{-i K}), \\
q_{12} &=&  t^{||}_1 e^{2 \pi i \varphi_1/{M}} e^{-i K} +t^{||}_2 e^{2 \pi i \varphi_2/{M}}\mu, \\
q_{21} &=& t^{||}_2 e^{2 \pi i \varphi_2/{M}}e^{-i K}+ t^{||}_1 e^{2 \pi i \varphi_1/{M}} \mu.
\end{eqnarray}
Other submatrices are null. Clearly, if $t^{\perp}_1=0$ and
$r_1=r_2$, the system reduces to the 1D case studied before.

If $\Delta r$ is small enough, the dominant elements of \eqref{HK}
are $H_{12}$ and $H_{13}$ and standard
perturbation theory can be applied. In first order approximation,
the spectrum of $H^{(K)}_{2}$ is given by the superimposition
of four sub-spectra. These sub-spectra are given by the spectrum of $-{\cal
F}(2\gamma, Q)/4$ shifted by four different values: $(1+\mu)
t^{\perp}_1$, $(1-\mu) t^{\perp}_1$, $(\mu-1) t^{\perp}_1$
and $(-1-\mu)t^{\perp}_1$, where
$Q=q_1+q_2+q_{12}+q_{21}$. According to this result, we expect
four bound states to exist but some of them can be hidden within the
continuum bands, as shown in  Fig.\ \ref{band} where just two of
them appear. Numerical tests
confirm that this first order perturbation result provides the
right width for the continuum bands, but fails to precisely
locate the bound states outside these bands. It also explains
the intra-band pattern observable in Fig.\ \ref{band}
in terms of overlapping bands of contiguous sub-spectra.

In the general case, it has not been possible to find an analytical
expression for the eigenvectors and eigenvalues of \eqref{HK}, but
they can be calculated numerically for different values of the
control parameters and even large values of the number of sites
$M$. In a similar way as in the previous section, we define
$E_\infty$ as the energy of the ground state when $M\rightarrow
\infty$ and $B=0$.  This quantity is a function of the distance
between both rings and its value can be determined numerically.

\subsection{Fast and slow AB oscillations}
\label{sec-2d-distance}

\begin{figure}[tbh]
\begin{center}
 \includegraphics[width=\figwidth]{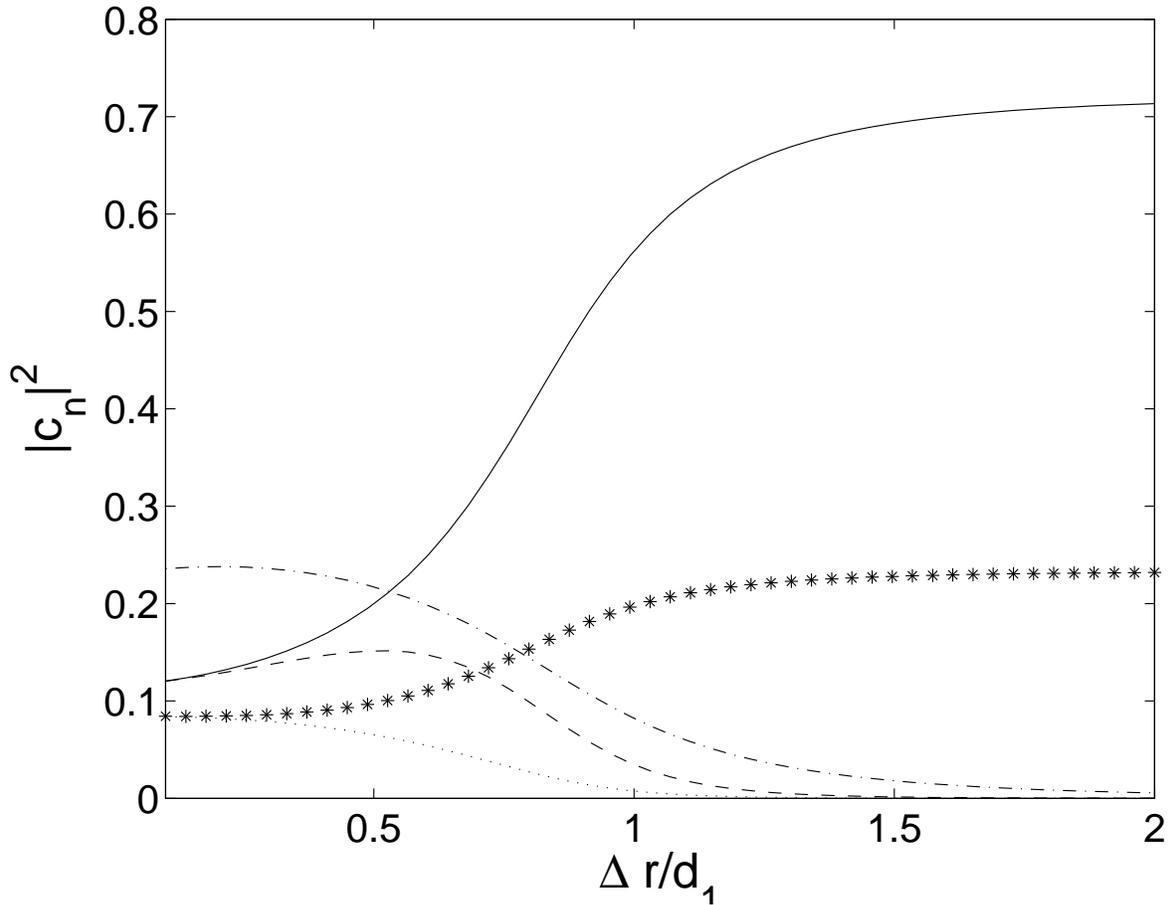}
 \caption{Main components of the wave function corresponding to the
   ground state as a function of the distance between rings $\Delta r$
   (in units of inner ring circumferential lattice spacing $d_1$). $\gamma=2.5$,
   $\epsilon=1$, $\mu=0.2$, ${M}=15$ and ${N}=2$. (---) two fermions in a
   bound state on the inner ring, (- -) two fermions in a bound state
   on the outer ring, ($\cdot -$) two fermions in a non-bound state
   in contiguous sites and an different rings, (*) two fermions in a
   non-bound state in contiguous sites and on the inner ring,
   ($\cdot\cdot\cdot$) two fermions in a non-bound state in
   contiguous sites and on the outer ring.}
  \label{comp}
\end{center}
\end{figure}

In general, for strong attraction $\gamma > 1$, as shown in Fig.\
\ref{comp}, the ground state is mainly a bound state, where
fermions are located together in the same location and on the same
ring. But in contrast to the 1D ring, we also find a
contribution of components corresponding to states where fermions
are in contiguous sites and on {\em different} rings as shown in
Fig.\ \ref{comp}.
Thus, the variation of the ground state energy with $\Phi/\Phi_0$
consists of a combination of periodic functions with different
frequencies. We find that the main contribution to the AB oscillations
corresponds to the components where the fermions are located around the
inner ring, with period $\Phi_1/\Phi_0$.
There exist also a further oscillation corresponding to the
contribution of the components where fermions are located around the
outer ring, with period $\Phi_2/\Phi_1$.
Due to the difference between hopping coefficients, a localization
effect around the inner ring takes place, and, as shown in
\cite{EilP04}, this effect is enhanced by the nonlinearity. Thus,
only when the distance between rings is very small, the
contribution of the periodic term with period $\Phi_2/\Phi_1$ is
(numerically) observable. In all cases, as in a 1D ring, the
contribution of these terms to the energy oscillations with the
magnetic flux decreases with the circumference of the ring.

There exists also a significant oscillation term coming from the
contribution of non-localized states, where the electron and the
hole are in contiguous sites but on different rings. These
oscillations persist when the circumference of the ring increases,
although their period increases too. We will refer to this
contribution as long-period oscillations. Essentially, the main
period of these oscillations is given by the phase difference
between the magnetic field term contribution in each ring.
Assuming that $\gamma=0$ and using Eq.\ (\ref{armonic}), this
period is given by $T=2M/(\Phi_2/\Phi_1-1)=2M/(r_2/r_1-1)=2 M r_1
/\Delta r$.
The amplitude of the long-period oscillations decreases when the
distance between the rings decreases. In cases where these
oscillations are dominant, corresponding to cases where the
distance between rings is small (or equivalently there is a large
hopping coefficients $t^{\perp}$ between rings), the oscillations
coming from components located around the same ring can be seen as
a small modulation that decreases with the circumference of the
ring.
We note that a model in which the hole is located more towards the
centre of the ring has been put forward \cite{GovUKW02} in the
context of the self-assembled nano-rings \cite{LorLGK00}.

In general, if the distance between the rings is small, the ground
state gives AB oscillations with period $\Phi_1/\Phi_0$. The first
excited state, where the main components corresponding to the
exciton are localized around the outer ring, oscillates with
period $\Phi_2/\Phi_0$. If the distance between the rings is
decreased, the hopping probability between different rings
increases and the bands overlap.  Then an additional oscillation
of the energy appears due to the non-localized components in
different rings.

\subsection{Flux dependence}
\label{sec-2d-flux}

\begin{figure}[tbh]
\begin{center}
\includegraphics[width=\figwidth]{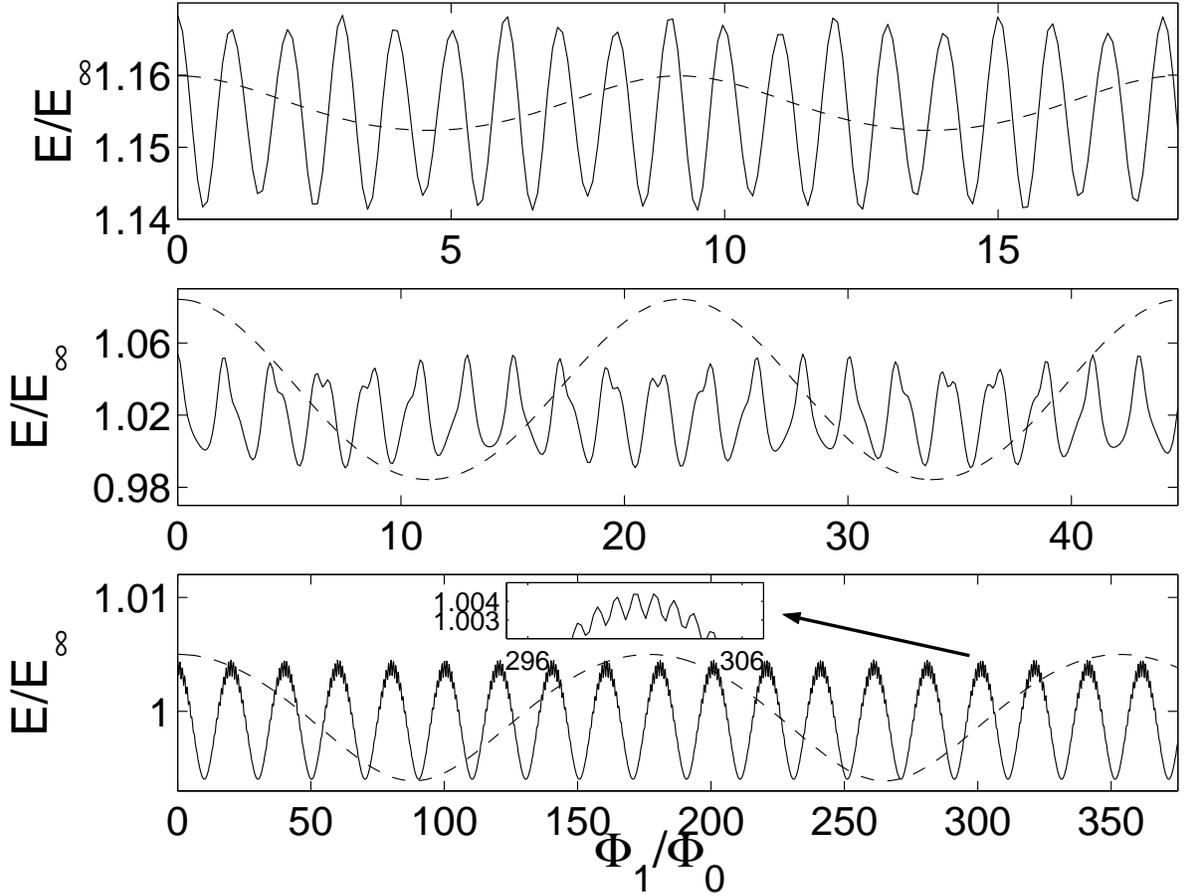}
\caption{Ground state energy oscillations of the exciton energy (in
  units of $E_\infty$) as a function of $\Phi_1/\Phi_0$ for several
  values of the distance between different rings $\Delta r=r_2-r_1$
  (in units of the circumferential lattice spacing $d_1$) and different values of
  the circumference of the inner ring $\rho_1$. $\gamma=2.5$,
  $\epsilon=1$, $\mu=0.2$ and ${N}=2$. $\Delta r=1.5$ (top), $\Delta
  r=0.7$ (centre) and $\Delta r=0.1$ (bottom).  Continuous lines
  correspond to ${M}=5$ ($\rho_1 \sim 5.34$) and dashed lines to
  ${M}=15$ ($\rho_1 \sim 15.11$).}
  \label{rem}
\end{center}
\end{figure}

In Fig.\ \ref{rem} we show the flux dependence of the ground state
energy for three different values of distance $\Delta r$ and two
different values of the circumference of the ring. In the top
figure we can see the small AB oscillations, corresponding to the
existence of the exciton state localized mainly around the inner
ring. If the circumference is small enough, the AB oscillations
are thus similar to the 1D case. If the circumference is larger,
the amplitude of the AB oscillations decreases and the long-period
oscillations become dominant.
If the distance between the rings is similar to the distance
between neighbouring sites and the circumference of the rings is
still small enough as shown in the central panel of Fig.\
\ref{rem}, the excitonic AB oscillations are completely mixed-in
with the the oscillations of similar frequencies coming from
non-bound components in different rings.
The bottom figure corresponds to the case where the two rings are
close to each other. If the circumference is small enough, the
excitonic AB oscillations appear as a slight modulation of the
non-excitonic oscillation due to the non-bound components in
different rings. If the circumference is large enough, only the
slow oscillations can be seen.

\subsection{Fourier analysis of AB oscillations}
\label{sec-2d-fourier}

\begin{figure}[tbh]
\begin{center}
\includegraphics[width=\figwidth]{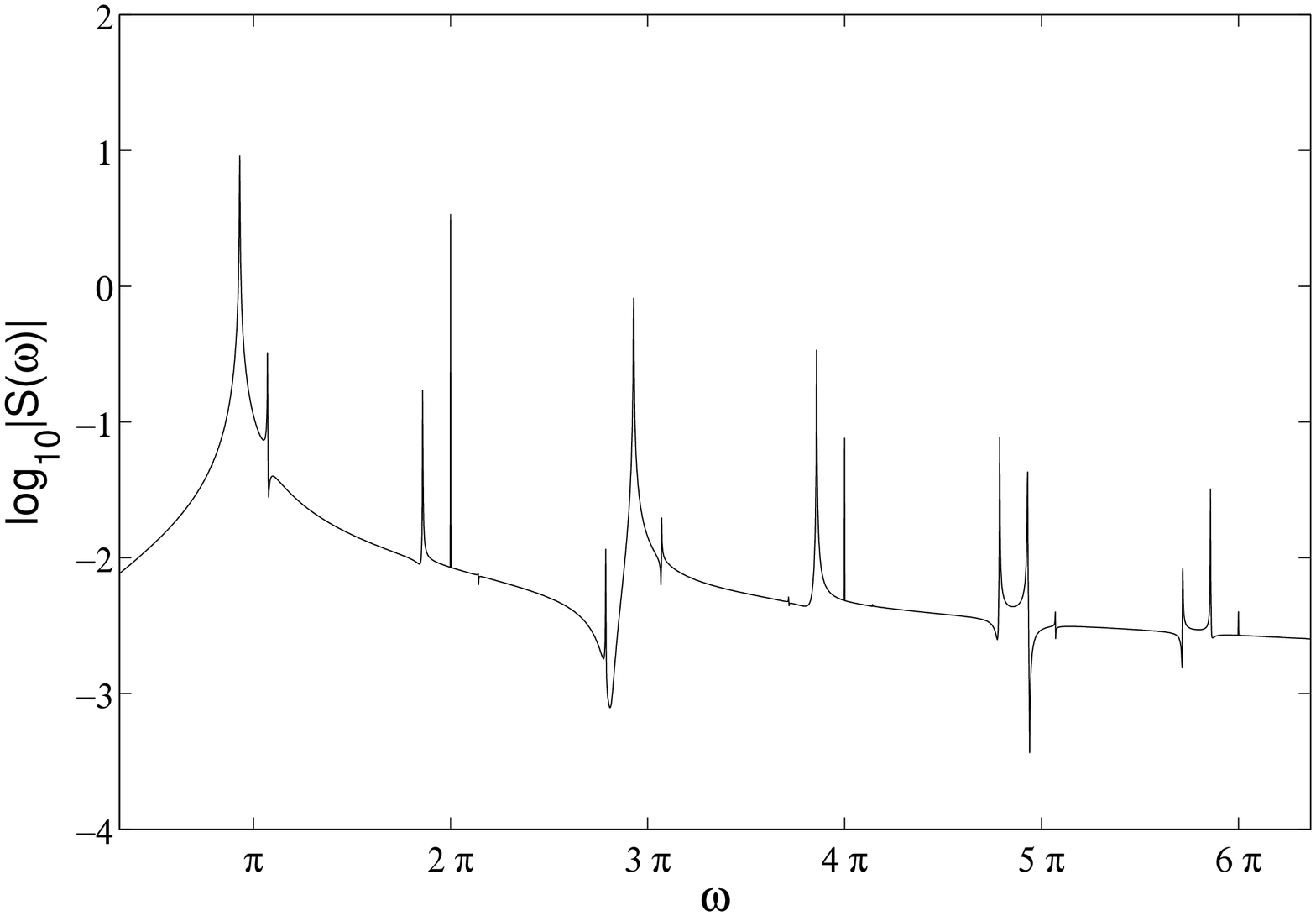}
\caption{Fourier spectrum of the ground state energy oscillations of
  the exciton energy (in units of $E_\infty$) as a function of magnetic
  flux $\Phi_1/\Phi_0$ for $\Delta r=0.7$ (in units of the reticular
  distance $d_1$) and ${M}=5$ ($\rho_1 \sim 5.34$). $\gamma=2.5$,
  $\epsilon=1$, $\mu=0.2$ and ${N}=2$. Note that resonances at the
  frequencies $\omega_1=2 \pi$, corresponding to AB oscillations in the
  inner ring, and $\omega_2=2.9$, corresponding to oscillations due to
  non-bound components in different rings (as well as their harmonics). All logarithms are
  common (base 10) logarithms.}
  \label{four_two}
\end{center}
\end{figure}

If we analyze the Fourier spectrum of the oscillations of ground state
energy as function of magnetic flux in cases where the AB oscillations
are not negligible (e.g., small ${M}$), and different values of distance
between rings $\Delta r$, as shown in Fig.\ \ref{four_two}, we clearly
detect the frequencies corresponding to AB oscillations in the inner
ring and the oscillations due to non-localized components in different
rings. A frequency corresponding to the contribution of components
around the outer ring can only be seen if the distance between rings is
very small ($\Delta r \simeq 0.1 d_1$) .

\subsection{Increasing the ring separation}
\label{sec-2d-ring_seperation}

\begin{figure}[tbh]
\begin{center}
\includegraphics[width=\figwidth]{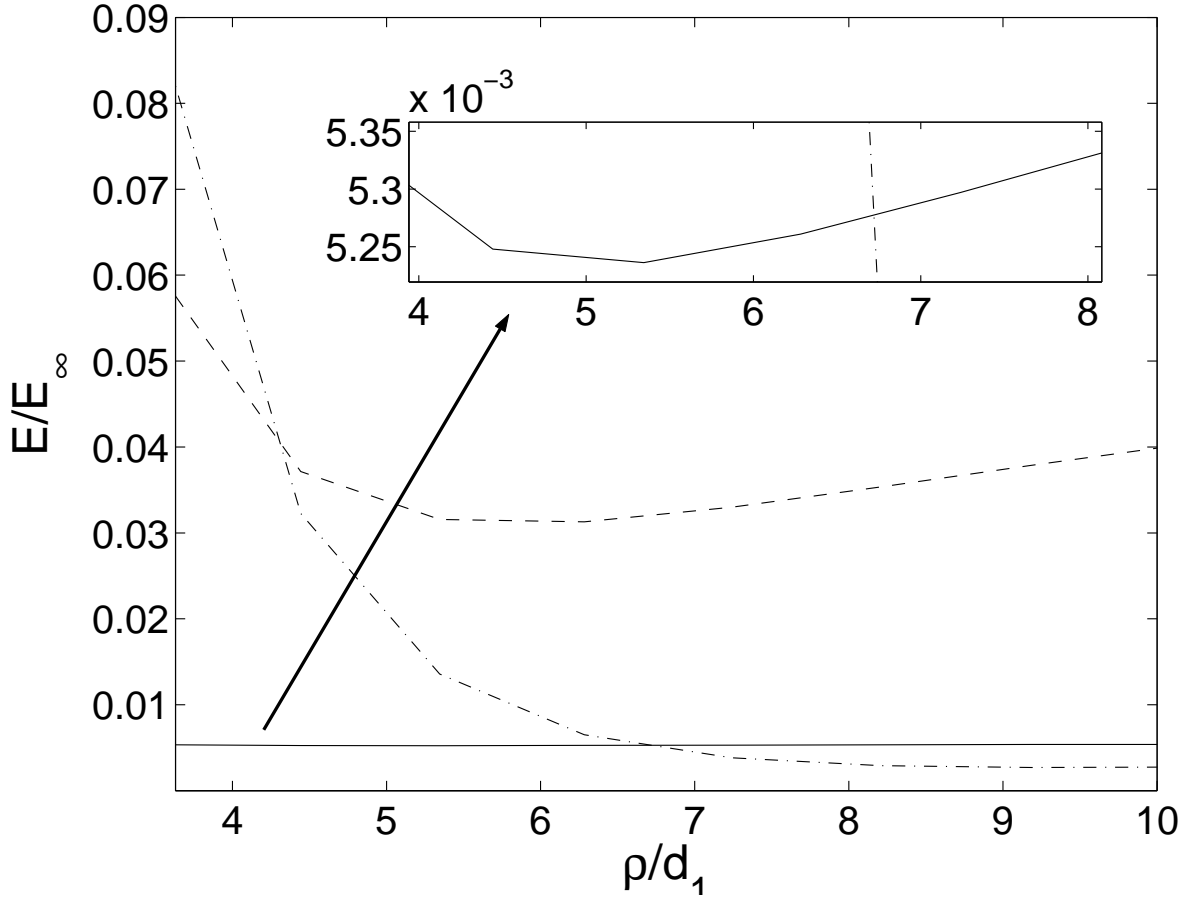}
\caption{Amplitude of the ground state energy oscillations (in units
  of $E_\infty$) as a function of the circumference of the ring $\rho$
  (in units of circumferential lattice spacing $d$), corresponding
  to different values of distance between rings $\Delta r=0.1$ (solid
  line), $\Delta r=0.7$ (dashed line) and $\Delta r=1.5$ (dot-dashed
  line).  $\gamma=2.5$ $\epsilon=1$, $\mu=0.2$ and ${N}=2$. The arrow
  indicates the region which has been shown in more detail in the inset.}
  \label{decay_two}
\end{center}
\end{figure}

If we study the decay of the amplitude $\Delta E$ of the ground state
energy oscillations as a function of the distance between different
rings, as shown in Fig.\ \ref{decay_two}, an approximately exponential
decay due to the AB effect can be detected, when the hopping between
different rings and the diameter of the ring are small.  In this case,
the typical decay length is similar to the 1D equivalent case. The most
interesting situation occurs when the hopping coefficient between
different rings is similar to the hopping coefficient along the inner
ring. If the circumference of the ring is small enough, AB oscillations
are significant and they decay with the radius of the ring.
Additionally, long--time oscillations appear whose amplitude {\em
  increases} slightly with the circumference (and thus the enclosed
flux) of the ring.  This leads to the non--monotonicity observed in
Fig.\ \ref{decay_two}.

\section{General case of $N$ annular rings}
\label{sec-general_case}

For arbitrary values of ${M}$ and ${N}$, we have not been able to
find a general expression for the Hamiltonian matrix.
Nevertheless, if ${M}$ and ${N}$ are small enough, it is possible
to calculate the eigenvalues and eigenvectors  by using algebraic
manipulation methods to construct an exact Hamiltonian matrix in
algebraic form and then use a numerical eigenvalue solver for each
parameter value.

\begin{figure}[tbh]
\begin{center}
 \includegraphics[width=\figwidth]{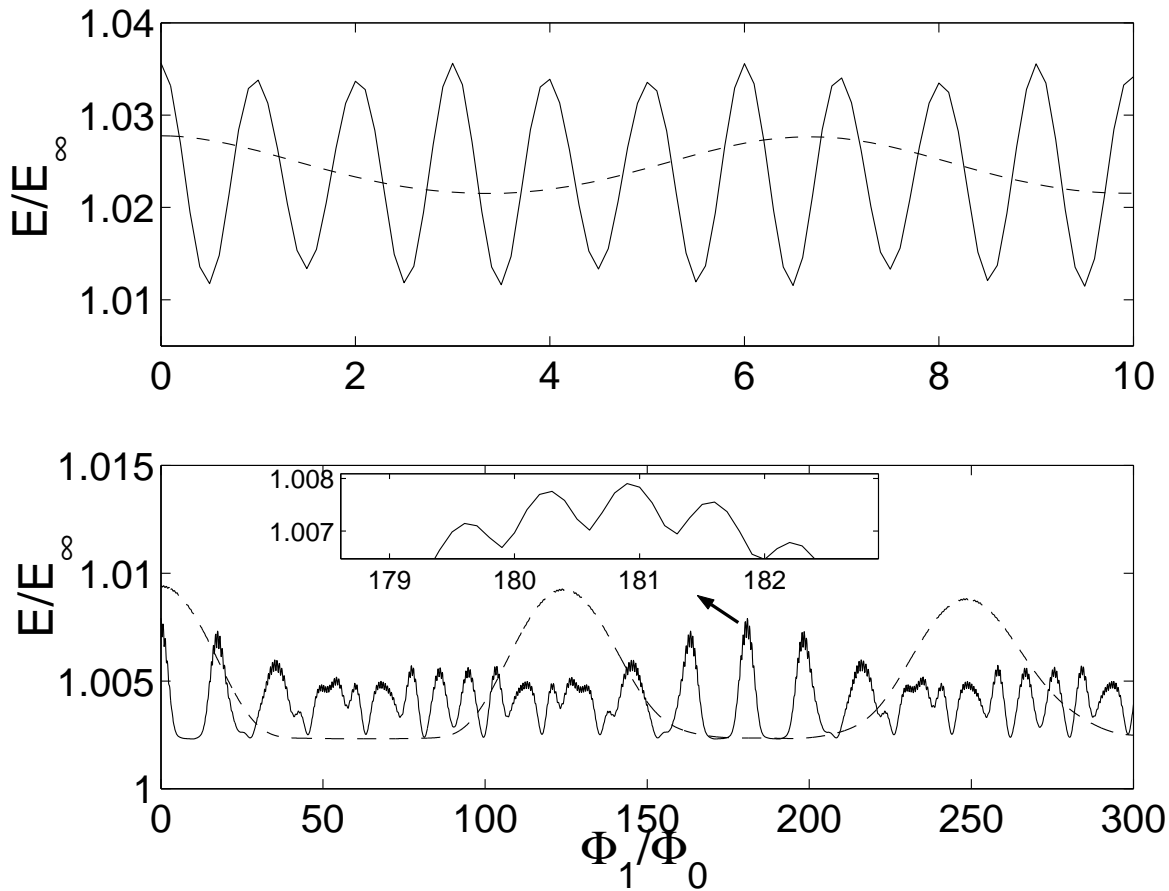}
  \caption{Aharonov-Bohm oscillations of the exciton energy
    (in units of $E_\infty$) as a function of $\Phi_1/\Phi_0$ for
    $N=5$ rings and
    several values of the distance between different rings $\Delta
    r=r_{i+1}-r_i$ and different values of the circumference of the
    inner ring $\rho_1$. $\gamma=2.5$, $d_1=1$, $\epsilon=1$, $\mu=0.2$,
    $k=0$. $\Delta r=1.5 d_1$ (top) and $\Delta r=0.1 d_1$
    (bottom).  Continuous lines correspond to ${M}=5$
    ($\rho_1=5.34$) and dashed lines to ${M}=13$ ($\rho_1=13.12$).}
  \label{energy_o_gen}
\end{center}
\end{figure}

As shown in Fig.\ \ref{energy_o_gen}, we have found essentially
the same phenomena as in the previous section. If the distance
between the rings is large enough, we can find AB oscillations due
to the existence of components corresponding to non-bound states
where the two fermions are in the inner ring. These oscillations
decrease very rapidly when the circumference of the rings
increases. When the distance between rings is small enough, there
are also significant long-period oscillations due to the
contribution of states corresponding to electrons and holes
located in neighbouring sites on different rings.

\begin{figure}[tbh]
\begin{center}
 \includegraphics[width=\figwidth]{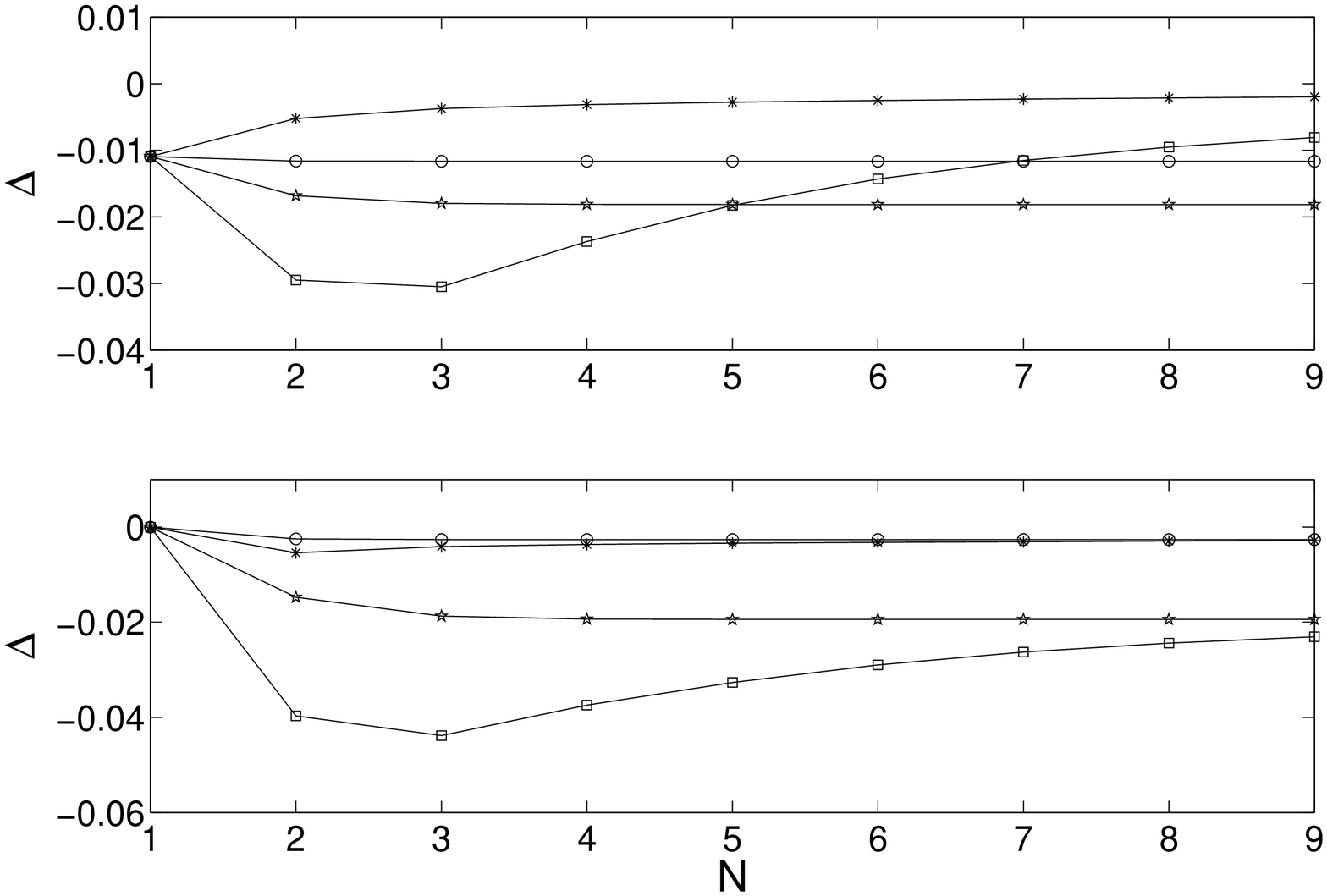}
 \caption{Oscillations of the energy of the ground state as function
   of the number of rings $N$. $\Delta=(E_{máx}-E_{min})/(2 E)$, where
   $E$ is the exciton energy for $B=0$.  $f=5$, corresponding to
   $\rho_1=5.34$ (top) and $f=11$, corresponding to $\rho_1=11.15$
   (bottom). (*) $\Delta r=0.1$, ($\square$) $\Delta r =0.7$,
   ($\star$) $\Delta r=1.0$ and ($\circ$) $\Delta r=1.5$.
   $\gamma=2.5$, $d_1=1$, $\epsilon=1$ and $\mu=0.2$.}
  \label{depg}
\end{center}
\end{figure}

If we study the amplitude of the ground state energy oscillations
as function of the number of rings $N$ as shown in Fig.\
\ref{depg}, we can observe that there exist a weak dependence with
the number of rings, but this dependence becomes negligible when
$N$ is more than a certain $N_c$, which in turn is a function of
$\Delta r$. If $\Delta r$ is large enough, $N_c \approx 1$. This
suggests that, except when the distance between rings is very
small, the dependence of the energy of the ground state with
magnetic field is essentially a 1D phenomenon.

\section{A system of stacked rings}
\label{sec-stack}

\begin{figure}[tbh]
\begin{center}
 \includegraphics[width=\figwidth]{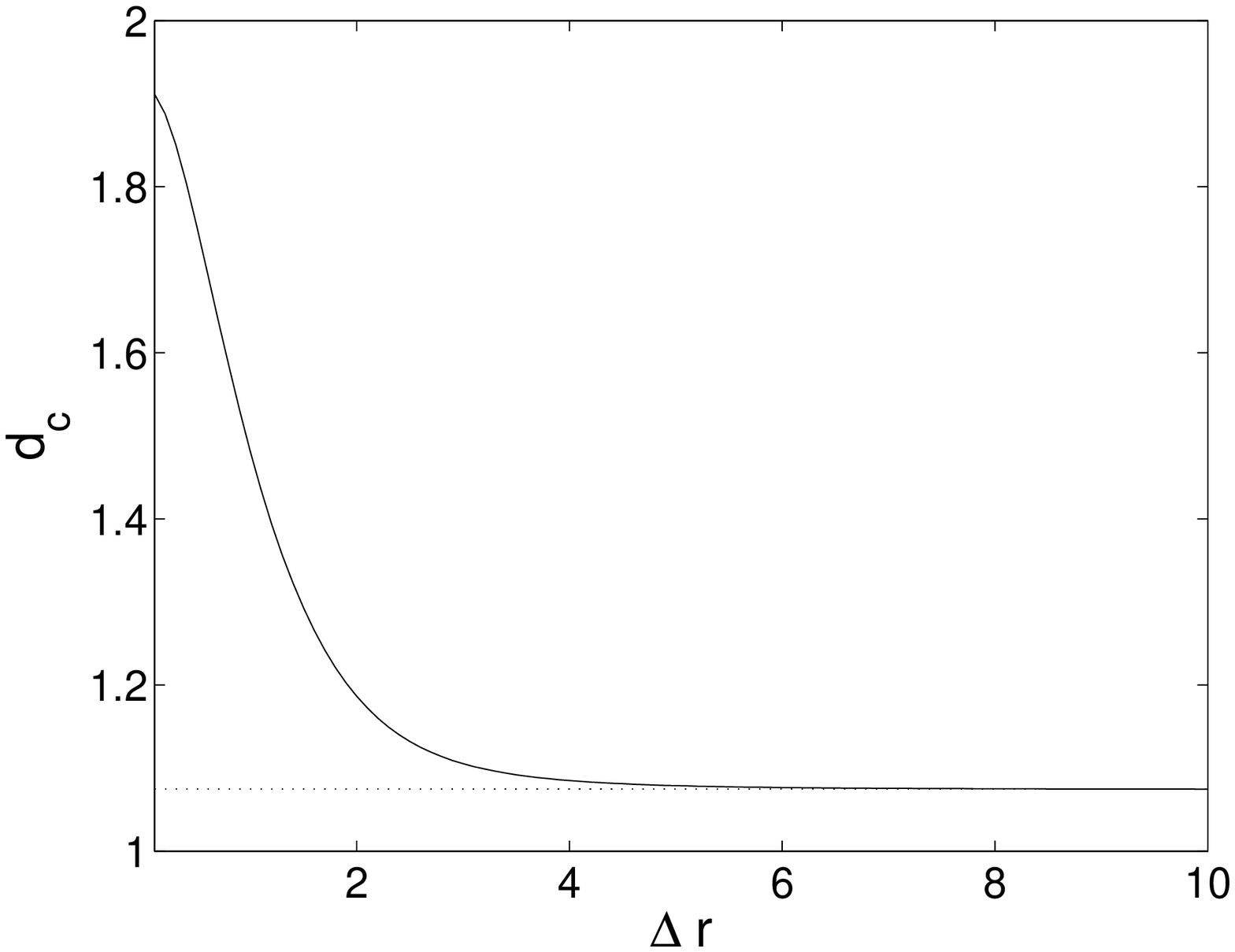}
 \caption{  \label{depg-stack}
   Decay length as function of the distance between rings corresponding
   to two stacked rings (continuous line) for $\gamma=2.5$ and reticular
   distance $d=1$. The dotted line corresponds to the 1D ring and has
   been calculated using Eq.\ \ref{EBS}.  }
\end{center}
\end{figure}

In the case of a {\em stacked} array of rings with {\em identical}
radius located along the magnetic field direction --- and thus all
of them threaded by the same magnetic flux --- our results are
similar to the single ring case. Only the oscillations of period
$T=\Phi_1/\Phi_0$ are observed.
If we further analyze the decay of the $\Delta E$ oscillations
with the circumference of the rings, we find that it is nearly
exponentially and the decay length can be studied.  As expected,
if $\Delta r$ is large, the stack behaviour is similar to isolated
rings.  But if the rings are close, the decay length {\em
increases} as shown in Fig.\ \ref{depg-stack}.
This suggests that the bound exciton states use the additional
rings as further channels along which to tunnel when prompted by
the increasing magnetic flux.

\section{Conclusions}
\label{sec-conclusions}

In this paper we have studied the existence of AB oscillations for an
exciton in a class of 2D annular models described by an attractive fermionic
Hubbard model. In all models, when the ring separation becomes
small, the 1D continuum results can be recovered in a limit, and the
structure of exact and approximate equations for the AB oscillations in
the discrete lattice model is similar to those for the continuum model.

Our results for the vertically-stacked ring system show that the
addition of further transport channels leads to an increase of the decay
length for the AB oscillations in the ground state. This effect
indicates a route to enhancing the possibility of experimentally
verifying the excitonic AB effect.

On the other hand, when additional transport channels are added
in-plane to give a disc-shaped, multi-ring structure as studied in
Sections \ref{sec-1d-intro} -- \ref{sec-general_case}, the
excitonic AB oscillations are quickly lost in the oscillations
which are due to the various additional flux periodicities in each
ring.  Already for the 2-ring case, our results show quite
complicated oscillation structures. Nevertheless, even in the
general case studied in Section \ref{sec-general_case}, remnants
of the original AB oscillations survive. These can be seen best
when analyzing the Fourier spectrum.

\acknowledgments
The authors are grateful for partial support under the LOCNET EU
network HPRN-CT-1999-00163. F.~Palmero thanks Heriot-Watt
University for hospitality, and the Secretar\'{\i}a de Estado de
Educaci\'on y Universidades (Spain) for financial support. We are
most grateful to our colleague Oliver Penrose for valuable
suggestions and discussions during the early stages of this
research

\appendix

\section{Eigenvalues and eigenvectors in a 1D ring}
\label{appendix}

We look for eigenvalues and eigenvectors of the matrix given by Eq.\
(\ref{ham_one}).
The corresponding eigenvalue equation is
\begin{equation} \label{eqHpsi}
\hat{H} |{\psi}\rangle = E |{\psi}\rangle,
\end{equation}
where $|{\psi}\rangle \equiv
\left(\psi_0,\psi_1,\cdots,\psi_{M-1}\right)^T\equiv\sum_{m=0}^{M-1}
\psi_m |m\rangle$ with suitable (i.e. normalized, complete and
orthogonal) atomic wave functions at each site $m$.
Due to the special circulant form of (\ref{ham_one}) (up to the
modified corner element $\gamma$) we may consider $|{\psi}\rangle$
as being $M$-periodic, that is $\psi_{m+M} = \psi_m$.

\subsection{Eigenvectors and eigenvalues}
\label{sec-eveceval}

The system of equations generated by (\ref{eqHpsi}) is
\begin{equation} \label{boundary}
\gamma \psi_{0} + q^* \psi_{1} + q \psi_{M-1} = -E \, \psi_{0} \,
,
\end{equation}
and
\begin{equation} \label{syst}
q \psi_{m-1} + q^* \psi_{m+1}  = -E \, \psi_{m}  \, ,
\end{equation}
with $m \in \{1,\cdots,M-1\}$ with periodic boundary conditions
$\psi_{m+M} = \psi_{m}$. In addition, we choose the phase of
$\psi_0$ to be zero, that is $\psi_0 \in \mathbb{R}$. This fixes
the overall phase factor up to which an eigenvector can be
determined. \noindent We now solve the difference equation
(\ref{syst}). Its characteristic equation reads
\begin{equation} \label{charac}
q^* r^2 + E r  + q = 0 \, ,
\end{equation}
whose solutions are
\begin{equation} \label{characsol}
r_{\pm} = \left( \varepsilon \pm \sqrt{\varepsilon^2 - 1}\right)
e^{i \theta} = e^{i (\theta \pm \nu)} \, ,
\end{equation}
where we have defined
\begin{equation} \label{defquant}
q = |q| e^{i \theta}\ \ \ ;\ \ \ \varepsilon = -\frac{E}{2
  |q|} := \cos{\nu}.
\end{equation}
Note that if $|\varepsilon| > 1$, $\nu$ becomes purely
imaginary. \noindent We can now write the solution of (\ref{syst})
as
\begin{equation} \label{solsyst}
\psi_{m} = \left[A \cos (\nu M) + B \sin (\nu m)\right] \, e^{i m
\theta}.
\end{equation}
It is obvious that $A=\psi_0$ in the equation above, and using the
periodicity condition $\psi_{0} = \psi_{M}$
\begin{equation} \label{cstB}
B  = \frac{\psi_0}{\sin (\nu M)} \left[ e^{- i M \theta}- \cos
(\nu M)\right].
\end{equation}
Thus
\begin{eqnarray}
\label{solsyst2} \psi_{m} &= &\frac{\psi_0}{\sin (\nu M)}
\left\{e^{i m \theta} \sin \left[(M-m) \nu\right]\right. \nonumber
\\
& & \left. \mbox{ } + e^{-i (M-m) \theta} \sin (\nu m)\right\}.
\end{eqnarray}
With the convention that $\psi_0 \in \mathbb{R}$, we obtain from
the latter equation that $\psi_{M-m} = \psi_m^*$. The
normalisation factor $\psi_0$ can be calculated exactly and reads
\begin{eqnarray} \label{normpsi0}
\lefteqn{\psi_{0} = } & & \\ & & \nonumber \frac{\sin
M\nu}{\left[M - \frac{\sin M\nu \cos M\nu}{\tan \nu} + \cos
M\theta\,\left(\frac{\sin M\nu}{\tan \nu}- M \cos M\nu  \right)
\right]^{1/2} }.
\end{eqnarray}

We now turn to the determination of the spectrum. Eigenvectors of
(\ref{ham_one}) have to verify the ``boundary condition''
(\ref{boundary}). Using the general expression derived in
(\ref{solsyst2}), this additional constraint results in
\begin{equation} \label{quantcond}
\frac{\tan (M \nu)}{\sin \nu} = \frac{2|q|}{\gamma} \left[1
  -\frac{\cos (M \theta)}{\cos (M \nu)}\right].
\end{equation}
Given that $\hat{H}$ is Hermitian, its eigenvalues are real. As we
have defined $E = -2 |q| \cos \nu$, we see that Eq.\
\eqref{quantcond} has to be solved for $\nu \in \mathbb{R}$, $\nu
\in i\mathbb{R}$ or $\nu \in \pi+i\mathbb{R}$ (the only distinct
possibilities for $E$ to be real). The solutions on the real axis
($\nu \in \mathbb{R}$) represent extended states whereas the
unique solution on the imaginary axis ($\nu \in i \mathbb{R}$ or
$\nu \in \pi+i \mathbb{R}$ ) if it exists, represents the bound
state. Once Eq.\ \eqref{quantcond} has been solved for $\nu$, the
spectrum is obtained from $E = -2 |q| \cos \nu$.

\subsection{Existence of a bound state}
\label{sec-bound}

We restrict hereafter to positive values of $\gamma$. It is
obvious that the case $\gamma<0$ can be handled from the latter as
$\hat{H}(-\gamma,q)=-\hat{H}(\gamma,e^{i\pi} q)$. For $\gamma
>0$ the eigenvector equation is obtained from \eqref{quantcond} with
$\nu= i \eta$, $\eta \in \mathbb{R}$. Then, \eqref{quantcond}
reads
\begin{equation} \label{quantcondBS}
\frac{\tanh (M \eta)}{\sinh \eta} = \frac{2|q|}{\gamma} \left[1
  -\frac{\cos (M \theta)}{\cosh (M \eta)}\right] \, .
\end{equation}
A necessary and sufficient condition for a bound state to exist is
given by Eq.\ (\ref{conditionb}).
To prove the result we first rewrite \eqref{quantcondBS} as
$f(\eta)=2|q|/\gamma$ where
$$f(\eta)=\frac{\sinh M\eta}{\sinh \eta \, (\cosh M\eta -\cos M\theta)}.$$
Differentiating $f$ with respect to $\eta$
we can show that, for $M \geq 1$ and $\eta>0$, $f'(\eta) <0$.



Now, $\lim_{\eta \rightarrow 0} f(\eta) = M/(1-\cos M\eta)$ and
$\lim_{\eta \rightarrow \infty} f(\eta) = 0$. Therefore, the
solution of $f(\eta)=2|q|/\gamma$ exists if $2|q|/\gamma \leq
M/(1-\cos M\eta)$ which establishes the result.

Note that when $\gamma = 2|q|/M(1-\cos M\theta)$, the solution
of \eqref{quantcondBS} is $\eta=0$. Then $\nu=0$ and
\eqref{solsyst2} shows that the bound state is not exponentially
localized because $\psi_m \propto e^{in\theta}(M-m+m
e^{-iM\theta})$. This limiting case cannot really be considered a
bound state, hence the strict inequality in \eqref{conditionb}.

\subsection{Approximation for large (but finite) number of sites M}
\label{sec-largeM}

In this case it is possible to derive explicit approximate
expressions for the eigenvalues both for extended and bound
states.

\subsubsection{Extended states ($\nu \in \mathbb{R}$)}

Let us rewrite (\ref{quantcond}) in the following form
\begin{equation} \label{appsolext}
\sin [M\nu - \varphi(\nu)] = - \sin \varphi(\nu) \, \cos(M\theta)
,
\end{equation}
where $\tan \varphi(\nu) = 2|q|\sin \nu/\gamma $. We can restrict
our investigation to the interval $\nu \in [0,\pi)$ within which
we will find $M-1$ eigenvalues or $M$ according to whether the
condition \eqref{conditionb} for the existence of a bound state is
satisfied or not. If $M$ is large, $\varphi(\nu)$ varies slowly as
compared to $M\nu$ and the zeroth order approximate solution for
(\ref{appsolext}) is
\begin{equation} \label{mu0}
\nu_l^{(0)} = \frac{l \pi}{M}\, ,\ \ \ l \in \{0,\cdots,M-1\}.
\end{equation}
Let us seek an approximate solution of the form $\nu_l =
\nu_l^{(0)}+ \nu_l^{(1)}$. Then the first correction is given by
\begin{eqnarray} \label{mu1}
\lefteqn{\nu_l^{(1)} = } & & \\ \nonumber & & \frac{1}{M}\left\{
\varphi\left(\nu_l^{(0)}\right) + \arcsin \left[
    (-1)^{l+1} \sin \varphi\left(\nu_l^{(0)}\right) \cos(M\theta)
    \right]\right\}.
\end{eqnarray}

\subsubsection{Bound state ($\nu \in i\mathbb{R}$)}

In this case, let $\nu = i\eta$. Then \eqref{quantcond} is given by
\eqref{enb}.
As $M \rightarrow \infty$, a first solution is
\begin{equation} \label{eta0}
\eta^{(0)} = {\rm arcsinh} \left( \frac{\gamma}{2|q|} \right),
\end{equation}
and the corresponding energy is
\begin{equation}
E^{(0)} = -2 |q| \cos \nu = -2 |q| \cosh \eta^{(0)} = -\sqrt{4
|q|^2 + \gamma^2},
\end{equation}
the solution for an infinite lattice. Notice that this value no longer
depends on $\theta$ (i.e. the magnetic field).
\noindent We can now use this last value to solve
(\ref{quantcondBS}) by a first Newton iteration for instance
\begin{equation} \label{eta1def}
\eta \simeq \eta^{(0)} + \eta^{(1)},
\end{equation}
where
$
\eta^{(1)}  = -\frac{f\left(\eta^{(0)}\right)}{f'\left(\eta^{(0)}\right)}
$ %
with
\begin{equation} f(x) = \frac{\tanh (M x)}{\sinh x} -
\frac{2|q|}{\gamma} \left[1
  -\frac{\cos (M \theta)}{\cosh (M x)}\right].
\end{equation}
As
$\eta^{(1)} \simeq -2 \tanh\eta^{(0)} \cos (M \theta) e^{-M \eta^{(0)}}$,
we finally obtain
\begin{eqnarray}
\lefteqn{E  = -2|q| \cosh \eta } \\ \nonumber &\simeq & -2|q|
\cosh \eta^{(0)} \left[1 + 2 \tanh2\eta^{(0)} \cos (M \theta) \,
e^{-M \eta^{(0)}} \right]
\nonumber \\
& = & \nonumber -\sqrt{4 |q|^2 + \gamma^2}\left[1 + \right.\\
\nonumber & & \mbox{ } \left.\frac{2\gamma^2} {4 |q|^2 + \gamma^2}
\cos (M \theta)\, e^{-M {\rm arcsinh} \left( \frac{\gamma}{2|q|}
\right)}\right].
\end{eqnarray}
The factor containing the magnetic field through $\cos (M \theta)$
decreases exponentially with the number of sites $M$, that is with
the radius of the ring (the lattice spacing is assumed to be
constant).


\end{document}